\documentclass{AIMS}
\usepackage{amsmath,amssymb}
\usepackage{paralist}
\usepackage{graphics} 
\usepackage{epsfig} 
\usepackage{graphicx}  
\usepackage{epstopdf} 
\usepackage[colorlinks=true]{hyperref}
\hypersetup{urlcolor=blue, citecolor=red}

\def\currentvolume{X}
 \def\currentissue{0}
  \def\currentyear{2018}
   \def\currentmonth{Month}
    \def\ppages{X--XX}
     \def\DOI{10.3934/xx.xx.xx.xx}

\theoremstyle{definition}

\usepackage{bm}
\newcommand{\re}{\mathrm{e}}

\newcommand{\rd}{\mathrm{d}}

\title[A $C$-integrable second sound equation] 
      {On a $C$-integrable equation for second sound propagation in heated dielectrics}

\author[I. C. Christov]{}

\subjclass{Primary: 35Q79, 80A20; Secondary: 35L20, 74J05.}
\keywords{Second sound, heat waves, $C$-integrability, dielectrics, radiation heat transfer, integral transforms.}

\email{christov@purdue.edu}

\begin{document}

\maketitle

\centerline{\scshape Ivan C. Christov}
\medskip
{\footnotesize
 \centerline{School of Mechanical Engineering, Purdue University}
 \centerline{West Lafayette, IN 47907, USA}
}

\bigskip


\begin{abstract}
An exactly solvable model in heat conduction is considered. The $C$-integrable ({\it i.e.}, change-of-variables-integrable) equation for second sound ({\it i.e.}, heat wave) propagation in a thin, rigid dielectric heat conductor uniformly heated on its lateral side by a surrounding medium under the Stefan--Boltzmann law is derived. A simple change-of-variables transformation is shown to exactly map the nonlinear governing partial differential equation to the classical linear telegrapher's equation. In a one-dimensional context, known integral-transform solutions of the latter are adapted to construct exact solutions relevant to heat transfer applications: (i) the initial-value problem on an infinite domain (the real line), and (ii) the initial-boundary-value problem on a semi-infinite domain (the half-line). Possible ``second law violations'' and restrictions on the $C$-transformation are noted for some sets of parameters.
\end{abstract}


\section{Introduction}

\begin{figure}[h]
\includegraphics[width=0.75\textwidth]{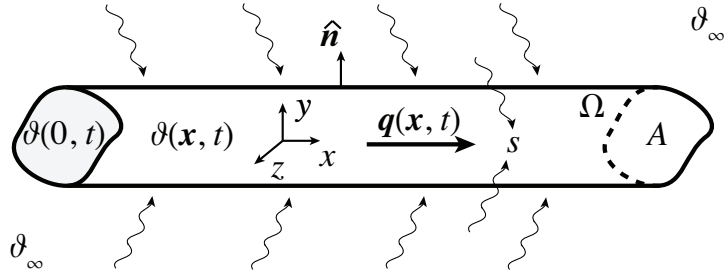}
\caption{Schematic of a thin, rigid rod of a dielectric material (contained within the domain $\Omega$ with boundary $\partial\Omega$ and unit surface normal $\bm{\hat{n}}$) subject to uniform heating/cooling of its lateral surface by its surroundings. The dielectric is long in the $x$-direction and thin in the cross-sectional $y$- and $z$-directions, so that heat conduction can be assumed to be unidirectional and radiation to be a volumetric source term in the energy equation. In this particular illustration, the temperature $\vartheta$ at one end ($x=0$) can be prescribed. The temperature in the surrounding medium ({\it i.e.}, in $\mathbb{R}^3\setminus\Omega$) is the constant $\vartheta_\infty$.}
\label{fig:schematic}
\end{figure}

Micro- and nano-scale heat transfer is a growing area of mechanical engineering with significant and timely technological applications. However, modeling heat transfer in such situations still remains challenging \cite{f14}. Yet, ``[i]nterface heat transfer is one of the major concerns in the design of microscale and nanoscale devices'' \cite{ghfx12}, specifically interface heat transfer between a dielectric and its surrounding medium ({\it e.g.}, a metal as investigated in \cite{ghfx12}). Here, following \cite{j15} (see also the time-fractional generalization of the latter \cite{g16}), we consider heat transfer in a thin, rigid rod of a dielectric material subject to uniform heating of its lateral surface by its surroundings, as shown schematically in Fig.~\ref{fig:schematic}. In this work, our goal is to show that such an inherently nonlinear problem can be transformed to a linear one, providing an exactly solvable heat transfer model.

In this context, the energy balance equation, in the absence of volumetric sources or sinks of heat, takes the form:
\begin{equation}
\frac{\partial e}{\partial t} + \bm{\nabla}\bm{\cdot}\bm{q} = 0,\qquad \bm{x}\in\Omega,
\label{eq:coe}
\end{equation}
where $\bm{q} = \bm{q}(\bm{x},t) = (q_x,q_y,q_z)$ and $e=e(\vartheta)$ denote the heat flux vector and the internal energy (per unit volume) in the material, respectively. At this stage, we consider $\bm{x} \equiv (x,y,z) \in \Omega\subset \mathbb{R}^3$, the boundary of the domain $\Omega$ being denoted by $\partial\Omega$ with unit surface normal $\bm{\hat{n}}$ (as shown in Fig.~\ref{fig:schematic}), and $t\ge0$. We assume that the heating/cooling of the conductor occurs via the Stefan--Boltzmann law \cite{cj59,mo13} of radiation/absorption of thermal energy across the dielectric's lateral surface into/from the surrounding medium. In other words, Eq.~\eqref{eq:coe} is supplemented by the boundary condition
\begin{equation}
\bm{q}\bm{\cdot}\bm{\hat{n}} = \sigma \epsilon \left(\vartheta^{4} - \vartheta_{\infty}^{4}\right),\qquad \bm{x}\in\partial\Omega,
\label{eq:SB}
\end{equation}
where $\vartheta=\vartheta(\bm{x},t)$ is the absolute temperature in the dielectric, $\vartheta_\infty$ is the surrounding medium's (constant) absolute temperature, the constant $\epsilon\in [0,1]$ is the emissivity of the surface of the dielectric, and $\sigma \approx 5.67 \times10^{-8}$ W/(m$^2$$\cdot$K$^4$) is the Stefan--Boltzmann constant \cite{mo13}. Equation~\eqref{eq:SB} is an approximation for the radiative heat transfer flux valid for convex surfaces placed into large isothermal environments (as in Fig.~\ref{fig:schematic}) \cite[p.~170]{mo13}. 

In Fig.~\ref{fig:schematic}, we have taken the dielectric to be (semi-)infinitely long and aligned with the $x$-axis, so that $\bm{\hat{n}}=(0,\hat{n}_y,\hat{n}_z)$ on its lateral surface. Additionally, the dielectric has a fixed cross-section $A$ of area $|A|$. In such a configuration, assuming a slender geometry, {\it i.e.}, the rod to be sufficiently long and thin so that lateral conduction in the cross-section establishes itself immediately, we may cross-sectionally average the energy Eq.~\eqref{eq:coe}:\footnote{Here, we have assumed, at least for the time being, the requisite smoothness of solutions to interchange the order of differentiation and integration.}
\begin{equation}
\frac{\partial \overline{e}}{\partial t} + \frac{\partial \overline{q_x}}{\partial x} + \frac{1}{|A|}\iint_A \bm{\nabla}_{\perp}\bm{\cdot}\bm{q} \, \mathrm{d}y\mathrm{d}z = 0, 
\label{eq:coe_avg_0}
\end{equation}
where we have introduced the notation $\bm{\nabla}_{\perp} := (\partial/\partial y, \partial/\partial z)$ for the gradient in the cross-sectional coordinates, and
\begin{equation}
\overline{(\,\cdot\,)} := \frac{1}{|A|}\iint_{A} (\,\cdot\,)\, \mathrm{d}y\mathrm{d}z.
\end{equation}
Now, using the divergence theorem, we rewrite Eq.~\eqref{eq:coe_avg_0} as
\begin{equation}
\frac{\partial \overline{e}}{\partial t} + \frac{\partial \overline{q_x}}{\partial x} + \frac{1}{|A|} \int_{\partial A} (q_y,q_z)\bm{\cdot}(\hat{n}_y,\hat{n}_z)|_{\partial\Omega} \,\mathrm{d}s = 0, 
\label{eq:coe_avg}
\end{equation}
where $\partial A$ is the boundary of the cross-sectional area $A$ with perimeter $|\partial A|$, and $\mathrm{d}s$ is the length element along it. Next, noting that $\bm{q}\bm{\cdot}\bm{\hat{n}} = (q_x,q_y,q_z)\bm{\cdot} (0,\hat{n}_y,\hat{n}_z) = (q_y,q_z)\bm{\cdot}(\hat{n}_y,\hat{n}_z)$ on ${\partial\Omega}$ and using the boundary condition from Eq.~\eqref{eq:SB}, Eq.~\eqref{eq:coe_avg} becomes
\begin{equation}
\frac{\partial \overline{e}}{\partial t} + \frac{\partial \overline{q_x}}{\partial x} =   - \frac{|\partial A|}{|A|} \frac{1}{|\partial A|} \int_{\partial A} \sigma \epsilon \left(\vartheta^{4} - \vartheta_{\infty}^{4}\right)\,\mathrm{d}s. 
\label{eq:coe_avg_00}
\end{equation}
Note that ${|\partial A|}/{|A|} \equiv 4/d_h$, where $d_h$ is, by definition, the \emph{hydraulic diameter} of the cross-section \cite[\S8.6]{blid11}, and this ratio is common in such cross-sectionally averaged one-dimensional heat conduction models (see also \cite[\S4.12]{cj59}). Finally, ``shrinking'' to a point, averages become point-values:
\begin{equation}
\frac{\partial e}{\partial t} + \frac{\partial q_x}{\partial x} =  - \frac{4}{d_h} \sigma \epsilon \left(\vartheta^{4} - \vartheta_{\infty}^{4}\right). 
\label{eq:coe_avg}
\end{equation}
In other words, for the long, slender geometry considered, the heating/cooling of the dielectric by its surroundings through radiation/absorption can be treated as a volumetric source term, \emph{instead of} a boundary condition. Note that Eq.~\eqref{eq:coe_avg} represents a \emph{nonlinear} partial differential equation (PDE).\footnote{A related nonlinear model was developed by Antaki \cite{a98}, in the context of reactive solids at high-temperature (constant $\tau$ and $K$), by taking the volumetric heating term to correspond to a single-step irreversible exothermic Arrhenius reaction. However, Antaki's model does not appear to be $C$-integrable.} Henceforth, we drop the $x$ subscript on $q$ without fear of confusion. For convenience, let us also re-normalize the Boltzmann constant as 
\begin{equation}
\hat{\sigma} := 4\sigma/d_h
\end{equation}
in the present one-dimensional context. 

We are interested in second sound propagation in this dielectric medium at either low temperatures (on the order of 1 K) or in the presence of ``high'' heat flux. For such applications, Chester \cite{c63} argued that the heat flux of certain dielectric solids should obey the Maxwell--Cattaneo (MC) \cite{m67,c48} constitutive relation/law (see also \cite{jp89,jp90,ios10,s11}):\footnote{The MC law appears under various names in the literature, including ``Cattaneo--Vernotte,'' due to Vernotte's seemingly parallel development of the mathematical model \cite{v58}. However, Catteneo promptly disputed said novelty claim \cite{c58}. In the related context of molecular diffusion and mass transfer, the history of the equivalent of Eq.~\eqref{eq:MC} is even more serpentine, going back at least two decades prior to Cattaneo's 1948 paper \cite{c48} to the Russian-language literature and the works of Fock and Davydov (see, {\it e.g.}, the discussion in \cite{b03,s18}).}
\begin{equation}
\tau(\vartheta) \frac{\partial q}{\partial t} + q = -K(\vartheta)\frac{\partial \vartheta}{\partial x},\label{eq:MC}
\end{equation}
where $K=K(\vartheta)>0$ is the thermal conductivity of the rigid conductor, and $\tau=\tau(\vartheta)$ is its thermal relaxation time. Note that for $\tau \equiv 0$, the MC law in Eq.~\eqref{eq:MC} reduces to Fourier's law of heat conduction, namely $q = -K(\vartheta)(\partial\vartheta/\partial x)$ (see also \cite{mr93,jcvl10}). The significance of the MC law, which has (since the works of Maxwell and Cattaneo) been put on solid ground through the theory of stochastic processes \cite{cm09} and using thermodynamic free energy and dissipation potentials \cite{os09} (see also \cite{mr93,jcvl10}), is that it allows for the propagation of \emph{heat waves}, also known as \emph{second sound} \cite{jp89,jp90,ios10,s11,c14}. For crystalline dielectric solids, Casimir--Debye theory predicts  (see, {\it e.g.}, \cite[\S8.1]{z62}) the following approximate forms of the conductivity and specific heat at constant volume: 
\begin{subequations}\begin{align}
K(\vartheta) &\approx K_\mathrm{R}\left(\frac{\vartheta}{\vartheta_\mathrm{R}}\right)^3,\label{eq:K}\\
c_v &\approx \beta \vartheta^3,\label{eq:cv}
\end{align}\label{eq:Kcv}\end{subequations}
both of which are valid for ``low'' temperatures, {\it e.g.}, $\vartheta\ll\min(\vartheta_\mathrm{D},\vartheta^*)$, where $\vartheta_\mathrm{D}$ is the Debye temperature and $\vartheta^*$ is the temperature at which the full conductivity curve might reach a maximum (see, {\it e.g.}, \cite[p.~292]{z62}). In Eq.~\eqref{eq:cv}, the constant $\beta (>0)$ is a property of the dielectric solid under consideration, which is in some ways similar to the Sommerfeld coefficient (per unit density) in metals \cite{D52}. Additionally, $K_\mathrm{R} = K(\vartheta_\mathrm{R})$ is the (reference) thermal conductivity of the solid at the (reference) temperature $\vartheta_\mathrm{R}$. According to Chester \cite[Eq.~(12)]{c63} (see also \cite[p.~186]{f72}), the thermal relaxation time for such a dielectric solid should be given by
\begin{equation}
\tau = \frac{3K}{\varsigma^2\varrho c_v},
\label{eq:taud}
\end{equation}
where $\varrho(>0)$ denotes the dielectric's constant (since it is a rigid solid) mass density, and $\varsigma$ represents the average value of the phonon speed in this medium.

In the present work, we derive the governing equation for the temperature field $\vartheta(\bm{x},t)$ in Section~\ref{sec:gov_eq}. In Section~\ref{sec:c-int}, we show that, in a one-dimensional (1D) context, the latter \emph{nonlinear} PDE can be \emph{exactly linearized} though a change-of-variables (``$C$'') transformation. Then, in Section~\ref{sec:exact_dwe}, we use exact solutions available for the latter linear PDE to obtain \emph{exact} solutions to the original \emph{nonlinear} boundary-value problem, while also providing graphical illustrations. Finally, in Section~\ref{sec:conc}, conclusions and avenues for future work are discussed.

\section{The governing equation}
\label{sec:gov_eq}

In this section, we seek to obtain the governing equation for second sound propagation in a uniformly heated long, thin and rigid dielectric rod. First, we use the equilibrium thermodynamic relation between \emph{specific} internal energy $u$ and temperature $\vartheta$ via the specific heat at constant volume $c_v$: $\mathrm{d}u = c_v \mathrm{d}\vartheta$. Noting that $u=e/\varrho$, from $\mathrm{d}e = \varrho c_v \mathrm{d}\vartheta$ and Eq.~\eqref{eq:Kcv}, it follows that
\begin{equation}
e(\vartheta) := \varrho\int c_v(\vartheta)\,\rd\vartheta = e_0 + \frac{\varrho\beta}{4}\vartheta^4,
\label{eq:e_theta}
\end{equation}
where $e_0$ is the internal energy (per unit volume) in some reference state, which can be taken to be the free vacuum so that $e_0=0$ without loss of generality. 
Substituting Eq.~\eqref{eq:e_theta} into Eq.~\eqref{eq:coe_avg}, the energy balance equation becomes
\begin{equation}
\left(\frac{\varrho\beta}{4}\right)\frac{\partial \vartheta^4}{\partial t} + \frac{\partial q}{\partial x} =  - \hat{\sigma} \epsilon \left(\vartheta^{4} - \vartheta_{\infty}^{4}\right).
\label{eq:coe2}
\end{equation}
Second, from Eqs.~\eqref{eq:taud} and \eqref{eq:Kcv}, we obtain
\begin{equation}
\tau(\vartheta) = \frac{3K_\mathrm{R}}{\varrho\varsigma^2 \vartheta_\mathrm{R}^3 \beta} = const.
\label{eq:taud_theta}
\end{equation}

Next, we substitute Eqs.~\eqref{eq:Kcv} and \eqref{eq:taud_theta} for the conductivity and relaxation time into the MC law for the heat flux, {\it i.e.}, Eq.~\eqref{eq:MC}, to obtain:
\begin{equation}
\frac{3K_\mathrm{R}}{\varrho\varsigma^2 \vartheta_\mathrm{R}^3 \beta} \frac{\partial q}{\partial t} + q = -K_\mathrm{R}\left(\frac{\vartheta}{\vartheta_r}\right)^3\frac{\partial\vartheta}{\partial x}.
\label{eq:MC2}
\end{equation}
To eliminate the heat flux $q$ between Eqs.~\eqref{eq:coe2} and \eqref{eq:MC2}, we apply $\partial/\partial t$ to Eq.~\eqref{eq:coe2}:
\begin{equation}
\left(\frac{\varrho\beta}{4}\right)\frac{\partial^2 \vartheta^4}{\partial t^2} + \frac{\partial^2 q}{\partial x\partial t} =  - \hat{\sigma} \epsilon \frac{\partial \vartheta^{4}}{\partial t}.
\label{eq:coe3}
\end{equation}
Then, we substitute for ${\partial q}/{\partial t}$ from Eq.~\eqref{eq:MC} into Eq.~\eqref{eq:coe3}:
\begin{equation}
\left(\frac{\varrho\beta}{4}\right)\frac{\partial^2 \vartheta^4}{\partial t^2} - \frac{\varrho\varsigma^2\beta \vartheta_\mathrm{R}^3 }{3K_\mathrm{R}} \frac{\partial q}{\partial x} -\frac{\varrho\varsigma^2 \beta}{3} \frac{\partial}{\partial x}\left[\frac{\partial}{\partial x}\left(\frac{\vartheta^4}{4}\right)\right] =  - \hat{\sigma} \epsilon \frac{\partial \vartheta^{4}}{\partial t}.
\label{eq:coe4}
\end{equation}
Next, we substitute for $\partial q/\partial x$ from Eq.~\eqref{eq:coe2} into Eq.~\eqref{eq:coe4} 
and simplify to arrive at the governing \emph{nonlinear} PDE for the temperature $\vartheta(x,t)$ field:
\begin{equation}
\frac{3K_\mathrm{R}}{4\varsigma^2 \vartheta_\mathrm{R}^3 \hat{\sigma}} \frac{\partial^2 \vartheta^4}{\partial t^2} + \frac{3K_\mathrm{R}}{4\varsigma^2 \vartheta_\mathrm{R}^3 \hat{\sigma}} \left(\frac{4\epsilon\hat{\sigma}}{\varrho\beta} + \frac{\varrho\varsigma^2 \beta\vartheta_\mathrm{R}^3}{3K_\mathrm{R}} \right)\frac{\partial \vartheta^{4}}{\partial t}  = \frac{K_\mathrm{R}}{4\vartheta_\mathrm{R}^3 \hat{\sigma}} \frac{\partial^2\vartheta^4}{\partial x^2} - \epsilon \left(\vartheta^{4} - \vartheta_{\infty}^{4}\right).
\label{eq:gov_eq_dim}
\end{equation}

For convenience, let us now introduce dimensionless variables through the following set of transformations:
\begin{equation}
\vartheta = \vartheta_\mathrm{c} \Theta,\qquad {x} = L_\mathrm{c} {X},\qquad t = \mathcal{T}_\mathrm{c} T,
\label{eq:nd_vars}
\end{equation}
where $\vartheta_\mathrm{c}$ and $L_\mathrm{c}$ are, respectively, characteristic temperature and length scales set by, {\it e.g.}, the initial or boundary conditions of a given problem or the geometry of a given domain. On the other hand, $\mathcal{T}_\mathrm{c}$ is the characteristic time scale emerging from the equation itself, namely
\begin{equation}
\mathcal{T}_\mathrm{c} = \sqrt{\frac{3K_\mathrm{R}}{4\varsigma^2 \vartheta_\mathrm{R}^3 \hat{\sigma}}}\, .
\end{equation}
Substituting the dimensionless variables from Eq.~\eqref{eq:nd_vars} into Eq.~\eqref{eq:gov_eq_dim} yields the dimensionless governing PDE:
\begin{equation}
\frac{\partial^2 \Theta^4}{\partial T^2} + \lambda_0 \frac{\partial \Theta^{4}}{\partial T}  = c_0^2 \frac{\partial^2 \Theta^4}{\partial X^2} - \epsilon (\Theta^{4} - \Theta_\mathrm{R}^4),
\label{eq:gov_eq_nd}
\end{equation}
where we have introduced three dimensionless numbers:
\begin{subequations}\begin{align}
\Theta_\mathrm{R} &:= \frac{\vartheta_\infty}{\vartheta_\mathrm{c}},\\
\lambda_0 &:= \sqrt{\frac{3K_\mathrm{R}}{4\varsigma^2 \vartheta_\mathrm{R}^3 \hat{\sigma}}}\left(\frac{4\epsilon\hat{\sigma}}{\varrho\beta} + \frac{\varrho\varsigma^2\beta \vartheta_\mathrm{R}^3}{3K_\mathrm{R}} \right),\\
c_0 &:= \sqrt{\frac{K_\mathrm{R}}{4\vartheta_\mathrm{R}^3 \hat{\sigma} L_\mathrm{c}^2}}.
\end{align}\end{subequations}
Note that $\lambda_0$ is a dimensionless effective \emph{inverse} thermal relaxation time, while $c_0$ is the dimensionless effective speed of second sound in the dielectric.

Obviously, if the length scale $L_\mathrm{c}$ is not set by the initial and/or boundary conditions, it can be taken to be $L_\mathrm{c} = \sqrt{K_\mathrm{R}/(4\vartheta_\mathrm{R}\hat{\sigma})}$ thus yielding $c_0=1$. In the present work, we consider only problems on the real line $X\in\Omega = (-\infty,+\infty)$ or the half-space $X\in\Omega =(0,+\infty)$ for which there is no natural length scale set by the initial or boundary conditions on these domains. Hence, we take $L_\mathrm{c} = \sqrt{K_\mathrm{R}/(4\vartheta_\mathrm{R}\hat{\sigma})}$ to be the ``natural'' length scale, and the speed of second sound $c_0$ is, thus, conveniently normalized to unity. 
Consequently, the final dimensionless form of the governing PDE \eqref{eq:gov_eq_nd} is
\begin{equation}
\frac{\partial^2 \Theta^4}{\partial T^2} + \lambda_0 \frac{\partial \Theta^{4}}{\partial T}  =  \frac{\partial^2 \Theta^4}{\partial X^2} - \epsilon (\Theta^{4} - \Theta_\mathrm{R}^4).
\label{eq:gov_eq_nd_1d}
\end{equation}
For the equivalent of Eq.~\eqref{eq:gov_eq_nd_1d} for the case of \emph{constant} specific heat $c_v$ and \emph{constant} thermal conductivity $K$, we refer the reader to \cite[p.~1020]{j03}. We now set out to \emph{exactly} solve Eq.~\eqref{eq:gov_eq_nd_1d} given certain initial and boundary conditions on the real line or on a half-space. Although it is, of course, evident that Eq.~\eqref{eq:gov_eq_nd_1d} is linear in $\Theta^4$, the question of whether simply writing down a general solution for $\Theta^4(X,T)$ (for given initial and boundary conditions) uniquely yields $\Theta(X,T)$ is far from obvious.

\section{$C$-integrability of the governing equation}
\label{sec:c-int}

In this section, we wish to establish whether and when a certain algebraic transformation can exactly linearize Eq.~\eqref{eq:gov_eq_nd_1d}. Here, ``$C$'' stands for ``change of variables'' \cite[p.~2]{c91}. Calogero \cite{c91,c17} has introduced this terminology to distinguish between nonlinear PDEs that can be linearized by a \emph{change-of-variables transformation} from $S$-integrable nonlinear PDEs that can be linearized via a spectral transform, {\it e.g.}, the \emph{inverse scattering transform} \cite{AS81}, which linearizes the celebrated Korteweg--de Vries equation \cite{kdv95,ggkm67}.

\subsection{Qualitative features and non-negativity of solutions}
\label{sec:qualitative}

First, note that the equilibrium solution, {\it i.e.}, the solution of Eq.~\eqref{eq:gov_eq_nd_1d} that is independent of ${X}$ and $T$, is simply $\Theta \equiv \Theta_\mathrm{R}$. Next, suppose a uniform initial condition is chosen such that 
\begin{equation}
\Theta^4|_{T=0} = \Theta_\mathrm{i}^4 \ne \Theta_\mathrm{R}^4 \quad\text{and}\quad \left.\frac{\partial\Theta^4}{\partial T}\right|_{T=0} = \hat\Theta_\mathrm{i}^{4} \qquad \forall {X}\in\Omega.
\label{eq:ode_ic}
\end{equation}
It follows from Eq.~\eqref{eq:gov_eq_nd_1d} that the temperature distribution equilibrates according to the ordinary differential equation (ODE):\footnote{Although analyzing a uniform state's time-evolution is an over-simplification over the original problem, there is some value in understanding the solutions to Eq.~\eqref{eq:1d_t_ode} because, as Jou {\it et al.}\ note ``[i]n continuous systems, the exponential decay observed in the discrete model would correspond to a diffusive perturbation, whereas an oscillation in [$\Theta$] would correspond to the propagation of a heat wave'' \cite[p.~46]{jcvl10}.}
\begin{equation}
\frac{\mathrm{d}^2 \Theta^4}{\mathrm{d} T^2} + \lambda_0 \frac{\mathrm{d} \Theta^{4}}{\mathrm{d} T}  = - \epsilon (\Theta^{4} - \Theta_\mathrm{R}^4).
\label{eq:1d_t_ode}
\end{equation}
Treating the latter as a linear ODE in the variable $\theta=\Theta^4$, the characteristic polynomial of the homogeneous ODE and its roots are easily found to be
\begin{equation}
r^2 + \lambda_0 r + \epsilon = 0 \qquad\Leftrightarrow\qquad r = r_\pm := \frac{1}{2}\left(-\lambda_0 \pm \sqrt{\lambda_0^2 - 4\epsilon}\right).
\end{equation}
Clearly, the ODE \eqref{eq:1d_t_ode} has \emph{oscillatory} solutions for $\lambda_0^2 < 4\epsilon$ because, then, $r_\pm \in \mathbb{C}$. 

We readily recognize these characteristic roots, and the ODE \eqref{eq:1d_t_ode} as the governing equation for damped, simple harmonic motion. Indeed, the case of $\lambda_0^2 < 4\epsilon$ is that of \emph{under-damped} simple harmonic motion. To determine non-negativity of solutions, we must first determine if the absolute minimum of $\Theta^4$, as a function of $T$ alone, can be $\le 0$. In the simple harmonic oscillator analogy, determining this condition can be thought of as a ``stopping problem'' for if and/or when the oscillator position reaches zero, having started from some non-zero position.

Upon finding the homogeneous and particular solutions to Eq.~\eqref{eq:1d_t_ode} and applying the initial conditions from Eq.~\eqref{eq:ode_ic}, we obtain
\begin{multline} 
\Theta^4(X,T) = \mathrm{e}^{-\lambda_0 T/2} \left\{\left[2\hat\Theta_\mathrm{i}^4 + (\Theta_\mathrm{i}^4 - \Theta_\mathrm{R}^4)\lambda_0\right] \frac{1}{\sqrt{4 \epsilon -\lambda_0^2}} \sin\left(\frac{1}{2} T \sqrt{4 \epsilon -\lambda_0^2}\right) \right. \\ \left. + \left(\Theta_\mathrm{i}^4 - \Theta_\mathrm{R}^4\right)  \cos\left(\frac{1}{2} T \sqrt{4 \epsilon -\lambda_0^2}\right)\right\} + \Theta_\mathrm{R}^4.
\label{eq:T4_osc_soln}
\end{multline}
Hence,
\begin{multline}
\frac{\mathrm{d}\Theta^4}{\mathrm{d}T} = \mathrm{e}^{-\lambda_0 T/2} \left\{ \left[2 \epsilon (\Theta_\mathrm{R} - \Theta_\mathrm{i}) - \lambda_0 \hat\Theta_\mathrm{i}^4 \right]\frac{1}{\sqrt{4 \epsilon -\lambda_0^2}}  \sin\left(\frac{1}{2} T \sqrt{4 \epsilon -\lambda_0^2}\right) \right. \\
\left. + \hat\Theta_\mathrm{i}^4 \cos\left(\frac{1}{2} T \sqrt{4 \epsilon -\lambda_0^2}\right)\right\},
\label{eq:gen_th_deriv}
\end{multline}
Although, in principle, one can now derive conditions, starting from Eq.~\eqref{eq:gen_th_deriv}, which guarantee that the first minimum of $\Theta^4$ remains non-negative, which by the exponential decay (in $T$) of the solution guarantees that $\Theta^4$ remains non-negative, these conditions are extremely involved. For the sake of brevity, and to illustrate just the main idea, let us consider the special case $\hat{\Theta}_\mathrm{i}^4=0$.

\begin{figure}
\centering
\includegraphics[width=0.6\textwidth]{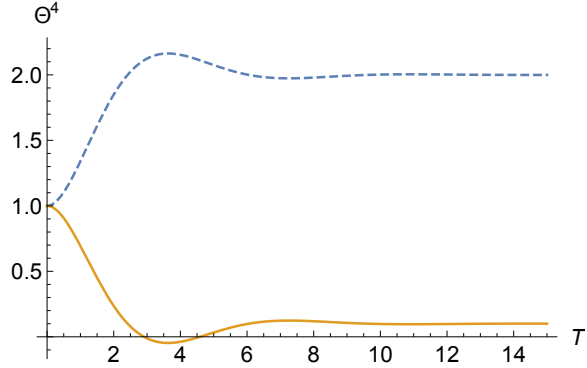}
\caption{Oscillatory behavior of solutions, given in Eq.~\eqref{eq:T4_osc_soln}, to the ODE \eqref{eq:1d_t_ode} for $\lambda_0=\epsilon=1 \Rightarrow \lambda_0^2 < 4\epsilon$. Solid curve correspond to $\Theta_\mathrm{i} = 1 > \Theta_\mathrm{R} = 0.1$, dashed curve corresponds to $\Theta_\mathrm{i} = 1 < \Theta_\mathrm{R} = 2$. For this choice of parameters, the solid curve's first minimum ``dips'' below $\Theta^4 =0$; thus, this solution is not strictly non-negative and $\Theta$ can become imaginary.}
\label{fig:Tosc}
\end{figure}

Now, from Eq.~\eqref{eq:gen_th_deriv} with $\hat{\Theta}_\mathrm{i}^4=0$, we find that the first extremum of $\Theta^4$ occurs at
\begin{equation}
T = T_1 := \frac{4 \pi}{\sqrt{4 \epsilon - \lambda_0^2}},
\end{equation}
If $\Theta_\mathrm{i} > \Theta_\mathrm{R}$, this extremum is a minimum (see solid curve in Fig.~\ref{fig:Tosc}), while for  $\Theta_\mathrm{i}<\Theta_\mathrm{R}$, it is a maximum (see dashed curve in Fig.~\ref{fig:Tosc}). For $\Theta_\mathrm{i} > \Theta_\mathrm{R}$, at $T=T_1$, we have
\begin{equation}
\Theta^4(T_1,X) = \exp\left(-\frac{2\pi\lambda_0}{\sqrt{4 \epsilon - \lambda_0^2}}\right) \left(\Theta_\mathrm{i}^4 - \Theta_\mathrm{R}^4\right) + \Theta_\mathrm{R}^4.
\end{equation}
Hence, $\Theta^4|_{T=T_1}<0$ if and only if
\begin{equation}
1 - \frac{\Theta_\mathrm{i}^4}{\Theta_\mathrm{R}^4} > \exp\left(\frac{2\pi\lambda_0}{\sqrt{4 \epsilon - \lambda_0^2}}\right).
\end{equation}
Equivalently, assuming that $\Theta_\mathrm{i}\ne\Theta_\mathrm{R}$, then strictly negative values of $\Theta^4$ ({\it i.e.},  strictly imaginary solutions for $\Theta$), occur if 
\begin{equation}
\lambda_0^2 < \frac{4 \epsilon}{1+4\pi^2/\ln\left|1 - {\Theta_\mathrm{i}^4}/{\Theta_\mathrm{R}^4}\right|^2} \qquad (\Theta_\mathrm{i}\ne\Theta_\mathrm{R}).
\label{eq:new_bound}
\end{equation}

Obviously, for $\Theta_\mathrm{i}=\Theta_\mathrm{R}$, $\Theta$ is simply the uniform steady state of Eq.~\eqref{eq:1d_t_ode} for all $T\ge0$, and there are no extrema in the solution profile. Note that the denominator in the bound in Eq.~\eqref{eq:new_bound} becomes singular for ${\Theta_\mathrm{i}^4}/{\Theta_\mathrm{R}^4} \approx 1$.\footnote{Although we use the approximate-equals sign, the critical value ${\Theta_\mathrm{i}^4}/{\Theta_\mathrm{R}^4} = 1 - \mathrm{e}^{-4 \pi ^2} $ is equal to $1$ within machine precision, {\it i.e.}, to sixteen decimals.} Hence, this special case in which the bound becomes unphysical ({\it i.e.}, $\lambda_0^2 < 0$) is not of significant physical interest as it is, for all practical purposes, the case of $\Theta_\mathrm{i}=\Theta_\mathrm{R}$. However, it can easily be shown from Eq.~\eqref{eq:new_bound} that, for $\Theta_\mathrm{i}^4/\Theta_\mathrm{R}^4 \in (0,2)$, the right-hand side of the bound in Eq.~\eqref{eq:new_bound} is \emph{negative}. Thus, since this is a bound on $\lambda_0^2$, then the \emph{only} choice of $\lambda_0$ that guarantees non-negativity of the solution in Eq.~\eqref{eq:T4_osc_soln}, for this range of $\Theta_\mathrm{i}^4/\Theta_\mathrm{R}^4$, is the trivial case $\lambda_0\equiv 0$ ({\it i.e.}, second-sound propagation does not occur). Hence, for clarity, we rewrite the bound, which is necessary to ensure non-negativity of $\Theta^4$ at its first minimum, as
\begin{equation}
\lambda_0^2 \begin{cases} \ge \displaystyle \frac{4 \epsilon}{1+4\pi^2/\ln\left|1 - {\Theta_\mathrm{i}^4}/{\Theta_\mathrm{R}^4}\right|^2} , &\quad \Theta_\mathrm{i}^4/\Theta_\mathrm{R}^4\not\in(0,2)\setminus\{1\},\\[3mm] =0, &\quad\text{else}.\end{cases}
\label{eq:new_bound2}
\end{equation}
For $\Theta_\mathrm{i}^4/\Theta_\mathrm{R}^4 \not\in (0,2)\setminus\{1\}$, the bound in Eq.~\eqref{eq:new_bound2} is \emph{sharper} than $\lambda_0^4 \ge 4\epsilon$, which means that there are non-negative oscillatory solutions to the ODE~\eqref{eq:1d_t_ode} for $\Theta^4$.

Clearly, the case $\lambda_0^2<4\epsilon$ [or $\lambda_0$ violating the bound in Eq.~\eqref{eq:new_bound2} for the special case of $(\partial\Theta^4/\partial T)_{T=0}=0$] is problematic. Of course, we are not the first to point out such a possibility; see, {\it e.g.}, \cite[p.~202]{jcvl10}. For example, Rubin \cite{r92} has argued that the possibility of negative temperature could be interpreted as a so-called ``second law violation.'' Such negative temperatures and second law violations typically arise if the thermal relaxation time is ``too large'' (compared to some appropriate time scale)\footnote{Others \cite[p.~228]{jcvl10} have formulated this condition as a restriction on the maximum magnitude of the heat flux at the initial instant of time.} and/or when a finite domain and reflections from boundaries are considered \cite{bl95}. In our context, $\lambda_0$ is a (dimensionless) \emph{inverse} thermal relaxation time, hence if we wish to avoid completely the issue of oscillatory solutions when $\lambda_0^2 < 4\epsilon$ can be interpreted as a \emph{upper} bound on the thermal relaxation, beyond which second law violations can occur. 

It is important to note that, in the present context, the emissivity $\epsilon$ of the dielectric plays a key role in setting the bound on $\lambda_0$ in Eq.~\eqref{eq:new_bound2}. Obviously, in the absence of heating/cooling of the dielectric ($\epsilon\equiv0$) in the above analysis, there are \emph{no} oscillatory solutions to Eq.~\eqref{eq:1d_t_ode} \emph{as long as} $(\partial\Theta/\partial T)|_{T=0}=0$. If $(\partial\Theta/\partial T)|_{T=0}\ne 0$ (but $\epsilon\equiv0$ still), then a restriction on $\lambda_0$ can be easily derived in terms of the initial conditions (which are, of course, related to the heat flux):
\begin{equation}
\lambda_0 \ge W_0\left[-\frac{(\partial\Theta^4/\partial T)|_{T=0}}{\Theta^4|_{T=0}}\right],
\label{eq:new_bound_eps0}
\end{equation}
where $W_0(\,\cdot\,)$ is the principal branch of the Lambert $W$-function \cite{cghjk96}, a special function with a surprising number of applications in the physical sciences (see, {\it e.g.}, \cite{pmj14}). As long as $\lambda_0$ satisfies Eq.~\eqref{eq:new_bound_eps0}, then solutions to the ODE \eqref{eq:1d_t_ode} with $\epsilon\equiv0$ remain non-negative for all $T\ge 0$. Note that, $(\partial\Theta^4/\partial T)|_{T=0}>0$ precludes the possibility of the first extremum being a minimum, hence non-negativity is guaranteed for all $\lambda_0>0$. Hence, for clarity, we may write the bound in Eq.~\eqref{eq:new_bound_eps0} as
\begin{equation}
\lambda_0 > \begin{cases} \displaystyle W_0\left[-\frac{(\partial\Theta^4/\partial T)|_{T=0}}{\Theta^4|_{T=0}}\right], &\quad(\partial\Theta^4/\partial T)|_{T=0}<0\\ 0, &\quad(\partial\Theta^4/\partial T)|_{T=0}>0.\end{cases}
\label{eq:new_bound_eps0_clear}
\end{equation}

Beyond restricting the value of $\lambda_0$ until no negative temperature arise, one can also consider modifications to the MC law proposed in \cite{cf082,r92,bl95,jcvl10} (amongst others) that do not exhibit such behaviors. The key modification in some of these theories \cite{cf082,bl95} is to introduce a \emph{non-equilibrium} relation of the form $e=e(\vartheta,\bm{q})$, in contrast to Eq.~\eqref{eq:e_theta}. Another modification could be to replace the MC law with one derived on the basis of Green--Naghdi theory \cite{gn95i,gn95ii,gn95iii}, as done in, e.g., \cite{js07,bsj08}. While this approach can be employed to study second-sound in rigid conductors, it is also not without criticism (see, e.g., the Mathematical Reviews entries for \cite{gn95i,gn95ii,gn95iii}). Nevertheless, thinking more broadly, perhaps ``second law violations'' are to be expected \emph{on average (in some probabilistic sense)} \cite{osm14,osr16} at the physical scales, temperatures and heat fluxes at which the MC law is believed to apply (and, hence, second sound exists).

\subsection{The $C$-transformation}
\label{sec:c-transform}

With the aforementioned issues in mind, let us now restrict to the case in which negative temperatures are not possible for the spatially-homogeneous temperature initial-value problem for Eq.~\eqref{eq:gov_eq_nd_1d}, specifically the bound $\lambda_0^2 \ge 4\epsilon$. The non-negativity of solutions of the spatially-homogeneous problem now guarantees that those solutions are bounded between $\max\{\Theta_\mathrm{i},\Theta_\mathrm{R}\}$ and $0$. However, these qualitative estimates do not necessarily apply to the the spatially-extended problem. Hence, in order to introduce a valid \emph{one-to-one} $C$-transformation, we consider two cases:
\begin{subequations}\begin{align}
\text{Case (I):}&\qquad 0 \le \Theta({X},0) \le \Theta_\mathrm{R} \qquad \forall {X}\in\Omega,\\
\text{Case (II):}&\qquad 0 \le \Theta_\mathrm{R} \le \Theta({X},0) \qquad \forall {X}\in\Omega. 
\end{align}\label{eq:cases}\end{subequations}
Then, the appropriate $C$-transformations are 
\begin{subequations}\begin{align}
\Theta^4-\Theta_\mathrm{R}^4 \mapsto \theta &\quad\Leftrightarrow\quad \Theta \mapsto \sqrt[4]{\theta + \Theta_\mathrm{R}^4} \qquad \text{[Case (I)]},\\
\Theta_\mathrm{R}^4-\Theta^4 \mapsto \theta &\quad\Leftrightarrow\quad \Theta \mapsto \sqrt[4]{\Theta_\mathrm{R}^4 - \theta} \qquad \text{[Case (II)]}.
\end{align}\label{eq:mappings}\end{subequations}
Since the dielectric conductor cannot be heated/cooled above/below the external medium's reference temperature $\Theta_\mathrm{R}$, we expect that the inequalities in Eq.~\eqref{eq:cases} hold true also for $T>0$. Hence, the $C$-transformations in Eq.~\eqref{eq:mappings} remain valid one-to-one mappings for $T>0$ as well.

Either of the mappings given in Eq.~\eqref{eq:mappings} renders Eq.~\eqref{eq:gov_eq_nd} as  \emph{$C$-integrable}. Now, it is easy to show that in either case (I) or case (II), the mappings introduced above transforms Eq.~\eqref{eq:gov_eq_nd_1d} into the \emph{telegrapher's equation} \cite{d74,b88,jp99}:
\begin{equation}
\frac{\partial^2 \theta}{\partial T^2} + \lambda_0 \frac{\partial \theta}{\partial T}  =  \frac{\partial^2 \theta}{\partial X^2} - \epsilon \theta.
\label{eq:gov_eq_lin_1d}
\end{equation}
In the absence of radiative heat transfer (equivalently, making the emissivity of the lateral surface vanish, {\it i.e.}, $\epsilon\equiv0$), we recognize that Eq.~\eqref{eq:gov_eq_lin_1d} becomes the well known \emph{damped wave equation} (DWE), which is discussed in the context of heat waves in \cite{c14}.

\section{Exact solutions to the $C$-linearized governing equation}
\label{sec:exact_dwe}

In this section, we adapt, to the present heat transfer context, known exact solutions to a standard initial-value problem (IVP) and a standard initial-boundary-value problem (IBVP) for  Eq.~\eqref{eq:gov_eq_lin_1d}. 

\subsection{d'Alembert problem on the real line}
\label{sec:sol_line}

In this subsection, let us consider the d'Alembert problem on the real line $\Omega = (-\infty,+\infty)$. We subject Eq.~\eqref{eq:gov_eq_lin_1d} to the following initial and asymptotic conditions:
\begin{equation}
\theta|_{T=0} = F(X),\qquad \left.\frac{\partial \theta}{\partial T}\right|_{T=0} = G(X),\qquad \lim_{|X|\to\infty} \theta(X,T) = 0 \;\; \forall T\ge 0, 
\end{equation}
where $F(\,\cdot\,)$ and $G(\,\cdot\,)$ are some arbitrary functions. The initial-value problem thus defined has an exact solution given in the textbook by Webster \cite[Ch.\ IV, Eq.~(146)]{w55} (see also the equivalent solution in Guenther and Lee's textbook \cite[\S5.5]{gl88}). 

To adapt the latter exact solution to the present context, first let us define $k:=\epsilon - \lambda_0^2/4$, which corresponds to the quantity $(ac-b^2)/a$ in \cite[Ch.\ IV, Eq.~(146)]{w55}. Webster only considers $(ac-b^2)/a<0$. Although we have restricted the value of $\lambda_0$ based on the considerations in Section~\ref{sec:qualitative}, we allow $k \le 0$. Hence, we must also consider the special case $(ac-b^2)/a=0$ explicitly, which is not difficult to do. Thus, we arrive at the exact solution
\begin{multline}
\theta(X,T) = \frac{1}{2}\re^{-\lambda_0 T/2}\Bigg\{F\left(X+T\right) + F\left(X-T\right) 
+ \int_{X-T}^{X+T} F(\eta) \mathcal{K}_1(X,\eta,T,k) \,\rd\eta \\
+ \int_{X-T}^{X+T}\left[G(\eta) + \frac{\lambda_0}{2}F(\eta)\right] \mathcal{K}_2(X,\eta,T,k) \,\rd\eta \Bigg\},
\label{eq:real_line_soln}
\end{multline}
where the integration kernels $\mathcal{K}_{1,2}$ are defined as
\begin{subequations}\begin{align}
\mathcal{K}_{1}(X,\eta,T,k) &:= \begin{cases} 0,\quad &k=0\\[1mm] -I_1\left( \sqrt{ |k| \left[T^2 - (X-\eta)^2\right]} \,\right) \displaystyle\frac{|k|T}{\sqrt{ |k| \left[T^2 - (X-\eta)^2\right]}},\quad &k<0 \end{cases},\\[2mm]
\mathcal{K}_{2}(X,\eta,T,k) &:= \begin{cases}1,\quad &k=0\\[1mm] I_0\left( \sqrt{ |k| \left[T^2 - (X-\eta)^2\right]} \,\right),\quad &k<0 \end{cases},
\end{align}\end{subequations}
and $I_\nu(\,\cdot\,)$ is the modified Bessel function of the first kind of order $\nu$. Notice that for $k=0$  (although $\lambda_0,\epsilon\ne0$) solutions to Eq.~\eqref{eq:gov_eq_lin_1d} on the real line {behave} similarly to solutions to the ``classical'' wave equation [{\it i.e.}, Eq.~\eqref{eq:gov_eq_lin_1d} with $\lambda_0\equiv0$ and $\epsilon\equiv0$], for which the d'Alembert solution is (see, {\it e.g.}, \cite[Ch.\ III, Eq.~(27)]{w55}):
\begin{equation}
\theta(X,T;\lambda_0=\epsilon=0) = \frac{1}{2}\Bigg[F\left(X+T\right) + F\left(X-T\right) + \int_{X-T}^{X+T} G(\eta)\,\rd\eta \Bigg].
\end{equation}
The solution in Eq.~\eqref{eq:real_line_soln} with $k=0$, nevertheless, retains the exponential damping in time, while also involving both initial conditions, {\it i.e.}, $F$ and $G$ into the integral term. The reader is referred to \cite{c14} for further discussion of wave solutions in heat transfer.

\begin{figure}
\centering
\includegraphics[width=0.75\textwidth]{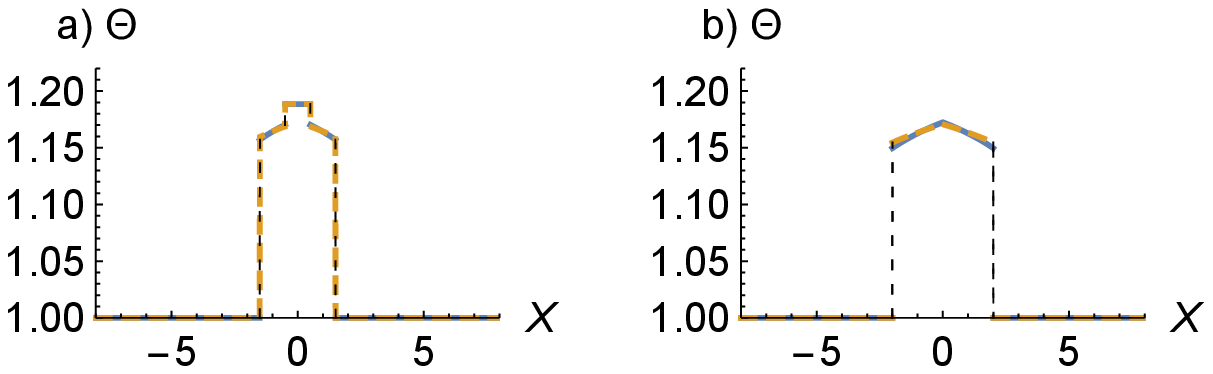}\\[3mm]
\includegraphics[width=0.75\textwidth]{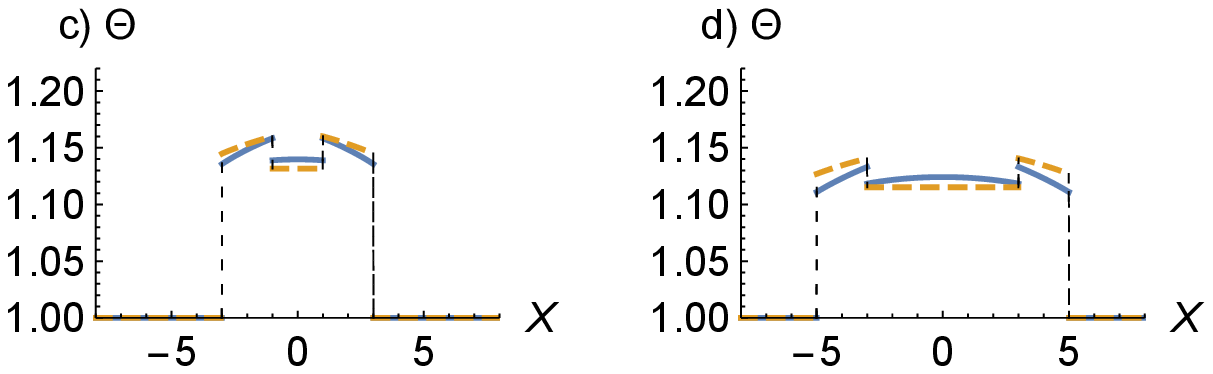}
\caption{Dimensionless temperature $\Theta$ profiles versus $X$ at different dimensionless times $T$ showing the relaxation of a unit pulse via the exact solution in Eq.~\eqref{eq:real_line_soln}. (a) $T=0.5$, (b) $T=1$, (c) $T=2$, (d) $T=4$. Here, $\epsilon=0.1$, $\lambda_0 = \tfrac{3}{2}\sqrt{4\epsilon}$ ($k<0$) for solid curves, while $\lambda_0 = \sqrt{4\epsilon}$ ($k=0$) for dashed curves.}
\label{fig:real_line_e01}
\end{figure}
\begin{figure}
\centering
\includegraphics[width=0.75\textwidth]{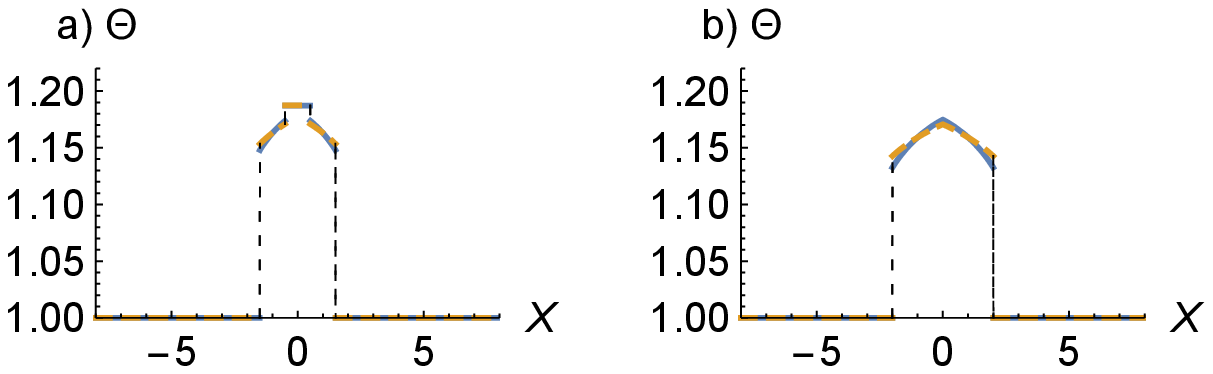}\\[3mm]
\includegraphics[width=0.75\textwidth]{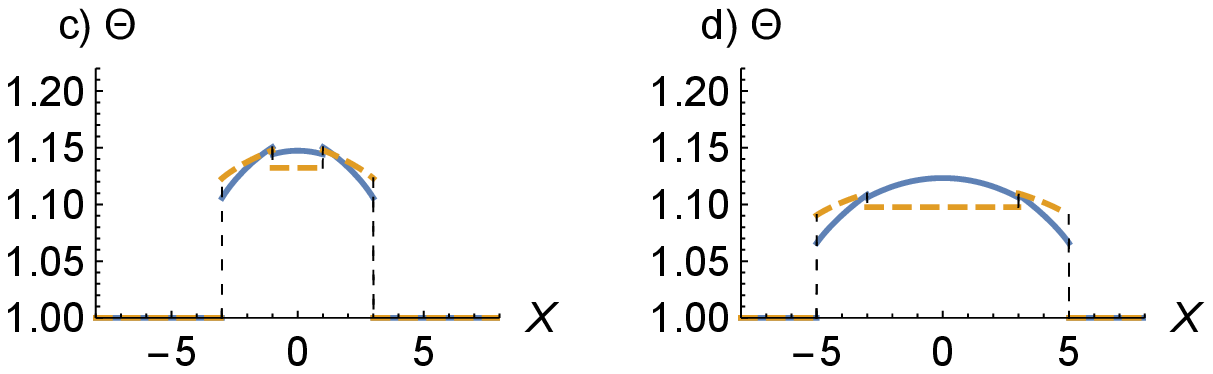}
\caption{Dimensionless temperature $\Theta$ profiles versus $X$ at different dimensionless times $T$ showing the relaxation of a unit pulse via the exact solution in Eq.~\eqref{eq:real_line_soln}. (a) $T=0.5$, (b) $T=1$, (c) $T=2$, (d) $T=4$. Here, $\epsilon=0.5$, $\lambda_0 = \tfrac{3}{2}\sqrt{4\epsilon}$ ($k<0$) for solid curves, while $\lambda_0 = \sqrt{4\epsilon}$ ($k=0$) for dashed curves.}
\label{fig:real_line_e05}
\end{figure}

Of interest is how an initial unit pulse (``box'') of heat, $F(X) = H(X+1)-H(X-1)$ and $G(X)=0$, relaxes in time. As $\Theta_\mathrm{R}$ has only a minor quantitative effect on the solutions depicted, we restrict our attention to $\Theta_\mathrm{R}=1$ without loss of generality. 
This solution is illustrated in Figs.~\ref{fig:real_line_e01} and \ref{fig:real_line_e05} for $\epsilon=0.1$ and $0.5$, respectively. Clearly, thermal shocks propagate outward from the left and right ($X=\pm1$) edges of the initial conditions along the space-time curves $X= \pm (1 + T)$, decaying in time. Additional shocks form and interact in the region in-between, {\it i.e.}, $-(1+T) < X < 1+T$. Notice, of course that the dashed curves in \emph{both} Figs.~\ref{fig:real_line_e01} ($\epsilon=0.1$) and \ref{fig:real_line_e05} ($\epsilon=0.5$) are \emph{qualitatively} similar with small quantitive difference in magnitude. This observation follows from the fact that, for $k=0$, the solution in Eq.~\eqref{eq:real_line_soln} no longer depends explicitly on the surface emissivity $\epsilon$. Nevertheless, we have chosen $\lambda_0 = \sqrt{4\epsilon}$, so the exponential damping of the solution changes with $\epsilon$.

\subsection{Signaling problem on a half-space}
\label{sec:sol_halfspace}

In this subsection, we consider the signaling problem on the half-space $\Omega = (0,+\infty)$. In the absence of radiative heat transfer (equivalently, making the emissivity of the later surface vanish, {\it i.e.}, $\epsilon\equiv0$), the exact solution of the signaling (also known as the thermal shock) problem for Eq.~\eqref{eq:gov_eq_lin_1d} was discussed in \cite[\S5]{j15}. We subject Eq.~\eqref{eq:gov_eq_lin_1d} to the following the initial, boundary and asymptotic conditions
\begin{equation}
\theta|_{T=0} = \left.\frac{\partial \theta}{\partial T}\right|_{T=0} = 0,\qquad \theta(0,T) = H(T)F(T),\qquad \lim_{X\to+\infty} \theta(X,T) =0 \;\;\forall T\ge 0,
\end{equation}
where $F(\,\cdot\,)$ is some arbitrary function, and $H(\,\cdot\,)$ is the Heaviside unit step function. The initial-boundary-value problem thus defined has an exact solution given by Jordan and Puri \cite{jp99}. 

To adapt the latter exact solution to the present context, we note that Jordan and Puri \cite{jp99} have considered all three cases $k\lesseqgtr0$ but we need only $k\le0$. Thus, we arrive at the exact solution
\begin{multline}
\theta(X,T) = H\left(T-X\right)\left[ \mathrm{e}^{-\lambda_0 X/2} F\left(T-X\right) + \int_{X}^T F\left(T-\eta\right) \mathcal{K}(X,\eta,|k|) \,\mathrm{d}\eta \right],
\label{eq:halfspace_soln}
\end{multline}
with $k:=\epsilon - \lambda_0^2/4$ as before and the integration kernel being defined as
\begin{equation}
\mathcal{K}(X,\eta,|k|) := \begin{cases} 0, \quad & k=0,\\[1mm] X\, \mathrm{e}^{-\lambda_0\eta/2} \displaystyle \frac{I_1\left[\sqrt{|k|(\eta^2-X^2)}\right]}{\sqrt{|k|(\eta^2-X^2)}}, \quad & k<0. \end{cases}
\end{equation}

\begin{figure}
\centering
\includegraphics[width=0.75\textwidth]{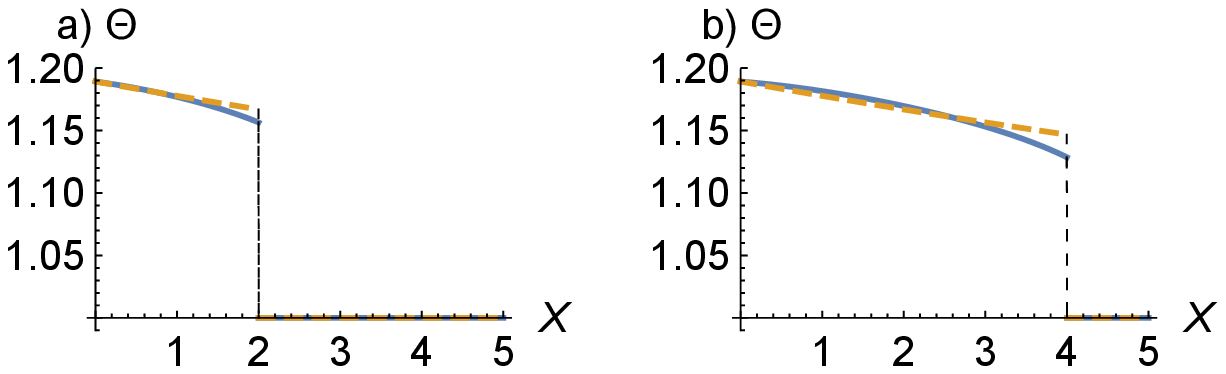}
\includegraphics[width=0.75\textwidth]{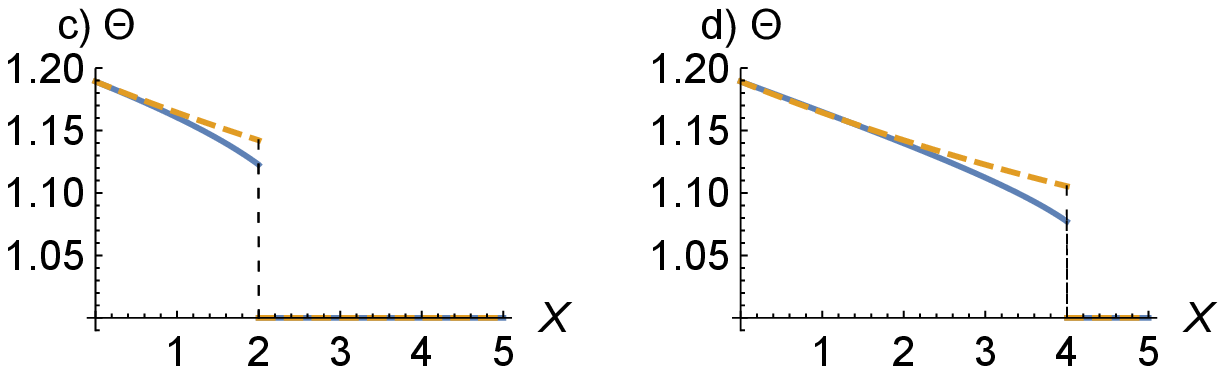}
\caption{Evolution of a (dimensionless) heat pulse $\Theta$ under the solution from Eq.~\eqref{eq:halfspace_soln}. (a,b) $\epsilon=0.1$, (c,d) $\epsilon=0.5$; (a,c) $T=2$, (b,d) $T=4$. In all panels $\lambda_0 = \tfrac{3}{2}\sqrt{4\epsilon}$ ($k<0$) for the solid curves, while $\lambda_0 = \sqrt{4\epsilon}$ ($k=0$) for the dashed curves.}
\label{fig:half_space}
\end{figure}

Of interest is the case of $F(T)=1$ $\forall T>0$, which corresponds to the so-called \emph{heat pulse} experiments \cite{ds93}, however, in the present case there is also lateral interfacial heating of the dielectric heat conductor. Again, as $\Theta_\mathrm{R}$ has only a minor quantitative effect on the solutions depicted, we restrict our attention to $\Theta_\mathrm{R}=1$ without loss of generality.  
This solution is illustrated in Fig.~\ref{fig:half_space}. Clearly, a shock propagates from the heated wall at $X=0$ into the medium along the space-time curve $X=T$, decaying in time. The amplitude of the shock, defined as $[\![\theta]\!](T) := \lim_{X\to T^-} \theta(X,T) - \lim_{X\to T^+} \theta(X,T)$, is easily found from Eq.~\eqref{eq:halfspace_soln} by taking $X\to T^\pm$ to obtain $[\![\theta]\!](T) = \mathrm{e}^{-\lambda_0 T/2}F(0)$. Now, using Eqs.~\eqref{eq:mappings},
\begin{subequations}\begin{align}
[\![\Theta]\!](T) = \sqrt[4]{\mathrm{e}^{-\lambda_0 T/2}F(0) + \Theta_\mathrm{R}^4} \qquad \text{[Case (I)]},\\
[\![\Theta]\!](T) = \sqrt[4]{\Theta_\mathrm{R}^4 - \mathrm{e}^{-\lambda_0 T/2}F(0)} \qquad \text{[Case (II)]}.
\end{align}\label{eq:jumps}\end{subequations}

Here, it is instructive to recall that Jordan and Puri note that ``[t]he case of $k<0$ is of much interest as it occurs in transmission line applications'' \cite[p.~1276]{jp99}. Jordan and Puri \cite{jp99} also show that necessarily $k<0$ when the telegrapher's equation is interpreted as a simplified model of both the current and voltage in a coaxial transmission line in the absence of external driving. Thus, it appears that this is also a case ($k<0$) of interest in low-temperature heat transfer in heated dielectrics. However, above we arrived at this condition \emph{not} by considering the physics of the problem \emph{but} by defining conditions on the validity of the $C$-transformation. Finally, we note that, in the context of the solution in Eq.~\eqref{eq:halfspace_soln}, the most pernicious feature of violating the bound on $\lambda_0$ ({\it i.e.}, takin $k>0$) is a change in the concavity of the profile behind the wave front ({\it i.e.}, for $X<T$), which leads to ``sagging'' and eventually negative values of $\theta$ (and, thus, imaginary values of $\Theta$) at some location behind the wavefront.

\section{Conclusion}
\label{sec:conc}

We considered an exactly solvable model in heat conduction, specifically the $C$-integrable ({\it i.e.}, change-of-variables-integrable) equation for second sound propagation in a thin, rigid dielectric heat conductor uniformly heated on its lateral side by a surrounding medium. The exact linearization of the governing PDE yielded the well known telegrapher's equation. The region of parameter space in which we expect the proposed $C$-transformation to be valid was motivated through a qualitative discussion on the non-negativity of solutions to the spatially-homogeneous problem. Adapting known transform solutions from the literature, we presented some relevant exact solutions to the heat transfer problem on an infinite domain (initial-value problem), and the heat pulse solution to the heat transfer problem on a semi-infinite domain (initial-boundary-value problem). Of course, due to the presence of jump discontinuities, the solutions discussed in Sections~\ref{sec:sol_line} and \ref{sec:sol_halfspace} must be regarded as \emph{weak}, satisfying the original PDE only in the sense of distributions.

Although we showed exact solutions are possible to what was, originally, a \emph{nonlinear} problem, it is also possible to ``attack'' the exactly $C$-linearized governing Eq.~\eqref{eq:gov_eq_lin_1d}  by either positivity-preserving \emph{nonstandard} finite-difference schemes, such as those constructed in \cite{j03,mj04}, or, if the governing equations are written as a first-order hyperbolic system, by Godunov-type shock-capturing schemes \cite{cj10}.

Note that if the heat conductor being considered is \emph{not} rigid, then the ``ordinary'' time derivative in the MC constitutive equation, {\it i.e.}, Eq.~\eqref{eq:MC}, must be replaced by an appropriate convected time rate as shown by C.~I.~Christov~\cite{CIC09}: the so-called ``Cattaneo--Christov'' heat flux law\footnote{However, the Cattaneo--Christov heat flux constitutive law and the work of Straughan {\it et al.}~\cite{s10,j14a,b16} should be \emph{clearly distinguished} from the  large and scientifically/mathematically questionable subset of the literature discussed in, {\it e.g.}, \cite{p17}.} \cite{s10}. Additional issues arise in fully deformable continua (such as dissipative solids \cite{m10}) and when mass transfer is also considered \cite{m13}, leading to the so-called ``Christov--Morro'' theory \cite{cst12,gs16}.

Finally, it would be of interest to develop sharper criteria for the non-negativity of solutions to the telegrapher equation, specifically for arbitrary spatially inhomogeneous initial conditions. Developing such a mathematical theory would extend the validity of the proposed $C$-transformation introduced in Section~\ref{sec:c-transform}, which maps the nonlinear Eq.~\eqref{eq:gov_eq_nd_1d} into the linear Eq.~\eqref{eq:gov_eq_lin_1d}, to a wider region of the $(\lambda_0,\epsilon)$ parameter space of the proposed heat transfer model.

\section*{Acknowledgements}
The author would like to thank Dr.\ P.\ M.\ Jordan and the guest editors for the invitation to participate in this special issue and for their efforts in organizing it. Helpful discussions with Dr.\ Jordan regarding Refs.~\cite{j03,j15} are also acknowledged. The author is also grateful to the anonymous reviewers for their helpful comments, which have improved the presentation and bibliography of this paper.


\providecommand{\href}[2]{#2}
\providecommand{\arxiv}[1]{\href{http://arxiv.org/abs/#1}{arXiv:#1}}
\providecommand{\url}[1]{\texttt{#1}}
\providecommand{\urlprefix}{URL }

\medskip
Received January 2018; revised April 2018.
\medskip

\end{document}